\documentclass{aa}
\usepackage{graphicx}
\usepackage{txfonts}
\usepackage{natbib}
\bibpunct{(}{)}{;}{a}{}{,}
\begin{document}
   \title{Observational evidence for the link between the variable
optical continuum and the jet of a radio-loud galaxy 3C\,390.3}

   \titlerunning{The link between the variable optical continuum and
   the relativistic jet}

   \authorrunning{Arshakian et al.}


   \author{Tigran G. Arshakian\inst{1},
	  Andrei P. Lobanov\inst{1},
	  Vahram H. Chavushyan\inst{2,3},
	  Alla I. Shapovalova\inst{4},
	  \and
	  J. Anton Zensus\inst{1}
          }

   \offprints{T.G. Arshakian}

   \institute{Max-Planck-Institut f\"ur Radioastronomie, Auf dem H\"ugel 69,
   53121 Bonn, Germany\\
   \email{tigar@mpifr-bonn.mpg.de}
         \and
	 Instituto Nacional de Astrof\'{\i}sica \'Optica y
         Electr\'onica, Apartado Postal 51 y 216, 72000 Puebla, Pue, M\'exico
	     \and
	     Instituto de Astronom\'{\i}a, UNAM, Apartado Postal 70-264,
	     04510 M\'exico D.F., M\'exico
	     \and
	     Special Astrophysical Observatory of the Russian AS, Nizhnij
	     Arkhyz, Karachaevo-Cherkesia 369167, Russia
             }

   \date{Received <date> / Accepted <date>}

   \abstract
{The ``central engine'' of AGN is thought to be powered by accretion on
a central nucleus believed to be a super-massive black hole. The
localization and exact mechanism of the energy release in AGN are
still not well understood.}
{We present observational evidence for the
link between variability of the radio emission of the compact jet,
optical and X-ray continua emission and ejections of new jet
components in the radio galaxy 3C\,390.3.} 
{The time delays between the light curves of the individual jet
 components and the light curve of the optical continuum are estimated
 by using minimization methods and the discret correlation function.}
{We find that the
variations of the optical continuum are correlated with radio emission
from a stationary feature in the jet. This correlation indicates that
the source of variable non-thermal continuum radiation is located in
the innermost part of the relativistic jet. }
{We suggest that the
continuum emission from the jet and counterjet ionizes material in a
subrelativistic outflow surrounding the jet, which results in a
formation of two conical regions with broad emission lines (in
addition to the conventional broad line region around the central
nucleus) at a distance $\ga$ 0.4 parsecs from the central engine.
Implications for modeling of the broad-line regions are discussed.}
  

   \maketitle
%

\section{Introduction}
The variable continuum flux in AGN, signaling the activity of the
central engine, is detected throughout the entire electromagnetic
spectrum, on time-scales from days to years
\citep{peterson02,zheng,wamsteker,shapo}. The continuum flux is
believed to be responsible for ionizing the cloud material in the
broad-line region (BLR). Localization of the source of the variable
continuum emission in AGN is therefore instrumental for understanding
the mechanism for release and transport of energy in active
galaxies. In radio-quiet AGN, representing about 90\,\% of the AGN
population, the presence of rapid X-ray flux variations and iron
emission line (Fe K$\alpha$) indicates that most of the soft X-ray
emission originates from the accretion disk \citep{mushotzky93}. In
radio-loud AGN, the activity of the central engine is accompanied by
highly-relativistic collimated outflows (jets) of plasma material
formed and accelerated in the vicinity of the black hole
\citep{ferrari98}. Inhomogeneities in the jet plasma appear as a
series of compact radio knots (jet components) observed on scales
ranging from several light weeks to about a kiloparsec
\citep{alef,kellermann04}. Continuum emission from the relativistic
jet dominates at all energies \citep[see][] {ulrich,worrall}, swamping
the X-ray emission associated with the accretion flow.  Hence, the
continuum variability in radio-loud AGN may be related to both the jet
and the instabilities of accretion flows \citep{mushotzky93,ulrich}
near the central engine.  The unification scheme \citep{urry95} of
radio-loud AGN suggests that the central powerful optical continuum
and broad emission lines are viewed directly in radio quasars and BL
Lacs, whereas in radio galaxies these can be hidden by an obscuring,
dusty torus, and therefore, the bulk of continuum and broad-line
emission in radio galaxies may be attributed to the relativistic jet
rather than the central engine. The presence of a positive correlation
between beamed synchrotron emission from the base of the jet and
optical nuclear emission in radio galaxies suggests that the optical
emission is non-thermal and may originate from a relativistic jet
\citep{chiaberge99,chiaberge02,hardcastle00}.  The detection of a
correlation between the radio and optical emission variability from
the nuclear region would be the most direct evidence of optical
continuum emission coming from the jet. No observational evidence has
yet been reported for a link between optical/UV continuum variability
and the radio jet in radio-loud AGN. In Section 2, we analyse the
structure and kinematics of the pc-scale jet in \object{3C 390.3} and
look for correlations between variable emission of jet components and
nuclear optical emission on scales less than one parsec. The structure
and emission mechanism of the sub-parsec-scale region around the
central nucleus is discussed in Section 3. In Section 4, we discuss
the results and draw conclusions.

\begin{figure}[t]
\resizebox{\hsize}{!}{\includegraphics[angle=-90]{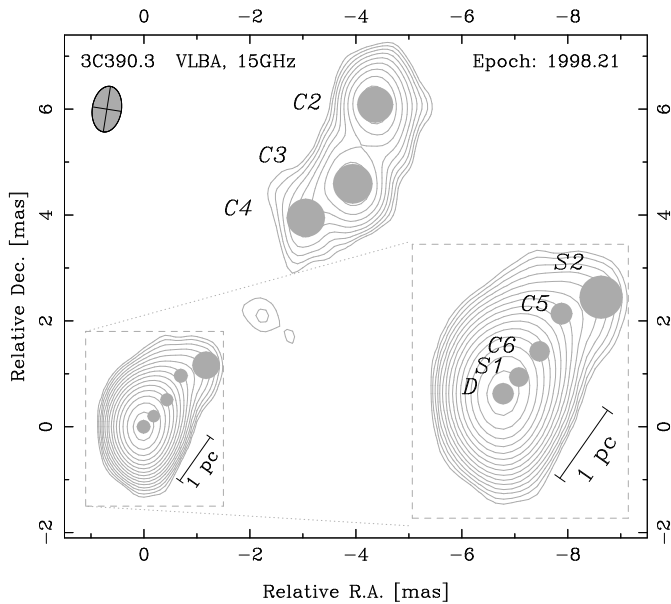}}
\caption{Radio structure of 3C 390.3 observed in 1998.21 with very
long baseline interferometry at 15\,GHz (2\,cm). Innermost fraction of
the jet is shown in the inset.  The resolving point-spread function
(beam) plotted in the upper left corner is $0.87 \times 0.55$ mas
oriented at an angle of 8.0 degrees (clockwise rotation).  The peak
flux density in the image is 190\,mJy/beam ($3.1\times 10^9$\,K) and
the rms noise is 0.2\,mJy/beam
($1\,\mathrm{Jy}=10^{-26}\,\mathrm{W}\,\mathrm{m}^{-2}\,\mathrm{Hz}^{-1}$).
The contours are drawn at $1,\,\sqrt{2},\,2\,...$ of the lowest
contour shown at 0.6\,mJy/beam. The structure observed is quantified
by a set of two-dimensional, circular Gaussian features (shaded
circles) obtained from fitting the visibility amplitudes and
phases \citep{pearson}. Similar fits have been obtained for ten
observations from the 15\,GHz VLBA survey database \citep{kellermann04},
for the purpose of cross-identifying and tracing different features in
the jet. The labels mark three stationary features (D, S1, and S2) and
a subset of moving components (C2--C6) identified in the jet. Note
that two more components, C7 and C8, have been first identified in the
jet in the VLBA images at later epochs (see Fig.~\ref{rfit}).}
\label{rmap}
\end{figure}

\section{The link between radio jet and optical continuum in the radio 
galaxy 3C 390.3} To search for a relation between variability of the
continuum flux and changes in the radio structure in AGN on scales of
$\sim 1$\,pc ($1\,\mathrm{pc} = 3.086\times 10^{16}$\,m), we combine
the results from monitoring of the radio-loud broad emission-line
galaxy 3C\,390.3 (redshift $z=0.0561$) in the optical
\citep{shapo,sergeev}, UV \citep{zheng}, and X-ray \citep{leighly}
regimes with ten very long baseline interferometry (VLBI) observations
of its radio emission at 15\,GHz \citep{kellermann04} made from 1992
to 2002 using the VLBA\footnote{Very Long Baseline Array of National
Radio Astronomy Observatory, Socorro, NM, USA}.

\subsection{Structure and kinematics of the parsec-scale jet} 
To parameterize the structure of the radio emission, we applied the
technique of modelfitting \citep{pearson} and fit interferometric
visibilities from each of the ten VLBA datasets by a set of circular
Gaussian components (Fig.~\ref{rmap} and Table~\ref{tbl-1} in the
Appendix) and used the positions and flux densities of these
components for tracing the evolution of the jet emission on angular
scales of $\sim 3$ milliarcseconds (mas). At the distance of
3C\,390.3, 1\,mas corresponds to a linear distance of 1.09\,pc
for the flat cosmology with the Hubble constant $H_0 =
70$\,km\,\,s$^{-1}$\,Mpc$^{-1}$ and the matter density
$\Omega_\mathrm{m} = 0.3$.

\begin{figure}
\resizebox{\hsize}{!}{\includegraphics{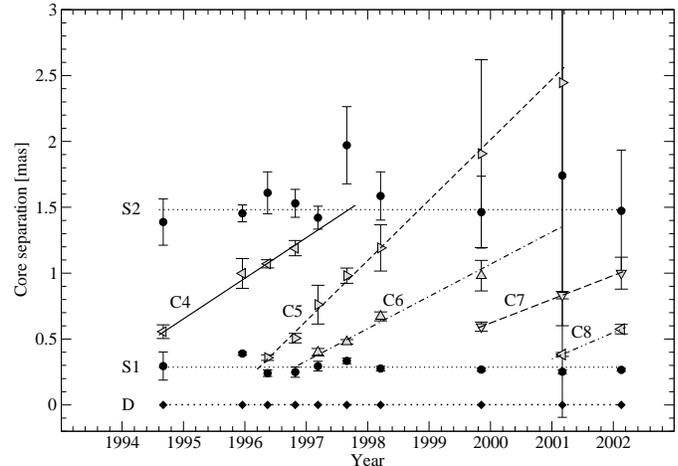}}
  \caption {Separation of the jet components relative to the
  stationary feature D (filled diamonds) for ten epochs of VLBI
  observations. Moving components are denoted by leftward triangles
  (C4), rightward triangles (C5), upward triangles (C6), downward
  triangles (C7), leftward triangles (C8), and stationary components
  S1 and S2 are marked by filled circles. The lines represent the
  best linear least-squares fits to the component separations.}
  \label{rfit}
\end{figure}

\begin{figure*}
\resizebox{\hsize}{!}{\includegraphics[angle=-90]{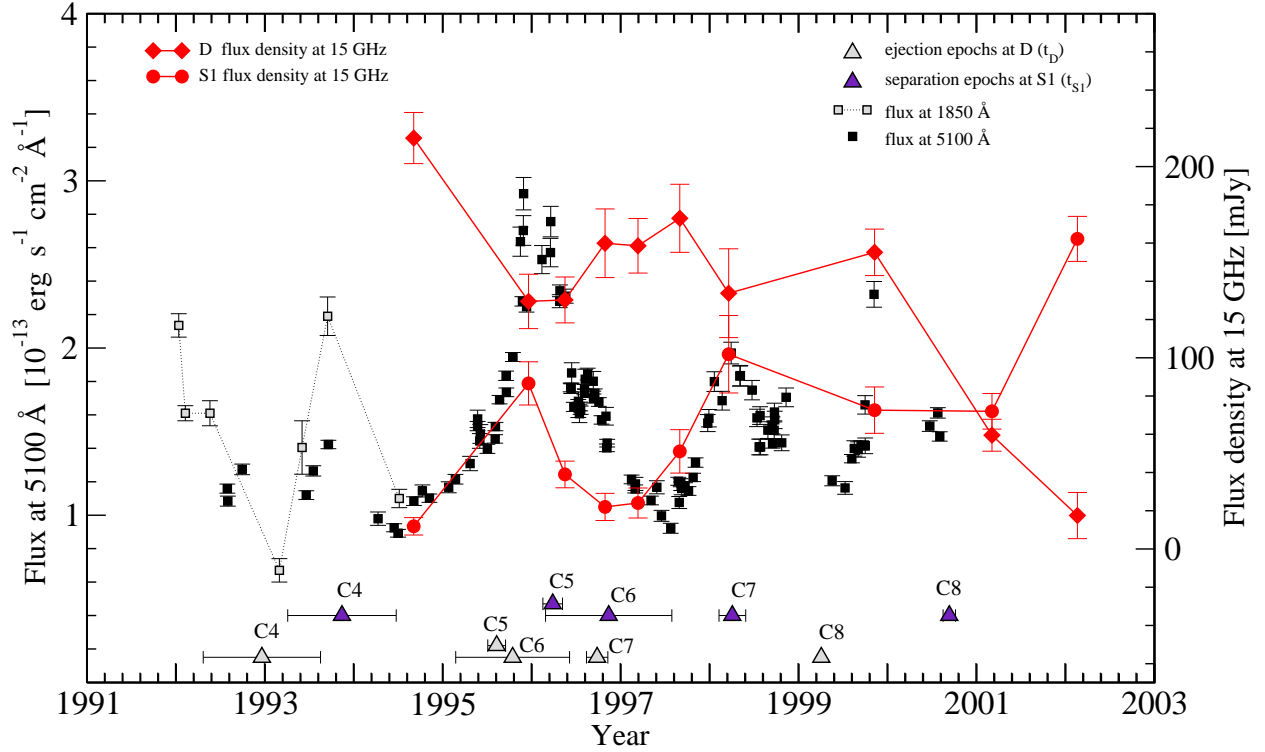}}
  \caption {The variations of the continuum fluxes of 3C 390.3 during
  the 1992 to 2002.2 time period. Optical continuum light curve at
  5100\AA\ \citep{shapo,sergeev} (black squares), UV continuum fluxes
  \citep{zheng} scaled by a factor of 50 (grey squares), and
  variations of radio flux density from D and S1 components (filled
  red diamonds and circles) are presented. The times of ejection,
  $t_\mathrm{D}$, of radio knots from D (presumed base of the jet) and
  the times of their separation, $t_\mathrm{S1}$, from the component
  S1 are marked by open triangles and filled triangles respectively.}
\label{flux-epoch}
\end{figure*}

From the component separations measured within 3\,mas from the feature
D at the narrow end of the jet (Fig.~\ref{rmap}), we identified five
moving components C4--C8 (in addition to the previously known
components C2 and C3; Alef et al. 1996) and two stationary features S1
and S2 separated from D by $0.28\pm0.03$\,mas and $1.50\pm0.12$\,mas,
respectively (Fig.~\ref{rfit}). The more distant stationary feature S2
can be explained by a small change in the direction of the flow that
causes the relativistic dimming of the radio emission
\citep{gomez97}. The presence of S1 may be related to both geometrical
and physical factors. Linear fits to the observed separations of the
component C4--C8 from the component D yield proper motions of
0.2--0.4\,mas/yr, which correspond to apparent speeds of
0.8--1.5\,$c$, where $c$ is the speed of light. Back-extrapolation of
the fits to C5 and C6 indicates that these two components may
originate from a single event. They could result from a moving
perturbation in the jet that creates a forward and a reverse shock
pair \citep{gomez97}. We used the linear fits to estimate, for each
moving component, the epoch, $t_\mathrm{D}$, at which it was ejected
from the component D and the epoch, $t_\mathrm{S1}$, when it passed
through the location of the stationary feature S1 (see
Figs.~\ref{rfit} and \ref{flux-epoch}).


\subsection{Correlations between variable emission of jet components and optical continuum}
In Fig.~\ref{flux-epoch}, the properties of the components D and S1
identified in the radio jet of 3C\,390.3 are compared with the
optical continuum emission at 5100\,\AA\ \citep{sergeev,shapo} and
1850\,\AA\ \citep{zheng}.
The radio fluxes from D ($f_{\rm D}$) and S1 ($f_{\rm S1}$) features
(Fig.~\ref{flux-epoch}) show an apparent anticorrelation during
1996.4-2002.2 time period, which indicates that there should be a time
delay between radio light curves. We should expect that the radio
flare propagates downstream of the jet, and hence the D component
should be the leading source of emission.

\begin{figure}
\resizebox{\hsize}{!}{\includegraphics{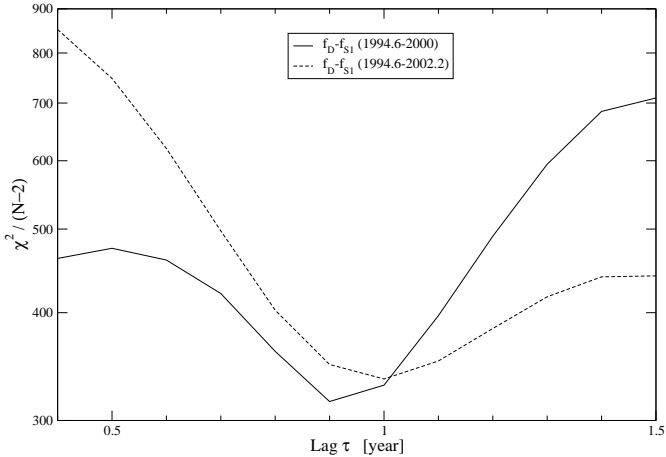}}
  \caption {The normalized $\chi^2$ statistics as a function of time
  delay between radio light-curves (ten data points between 1994.6 and
  2002.2) of D and S1 components. This dependence is presented for the
  entire radio data (dashed line) and for a subsample of eight radio
  epochs between 1994.6 and 2000 (solid line). The minimum corresponds
  to the best-fit delay. The best-fit delays are found in the range
  between 0.9 years to 1 year.}  \label{D-S1}
\end{figure}

\begin{figure}
\resizebox{\hsize}{!}{\includegraphics{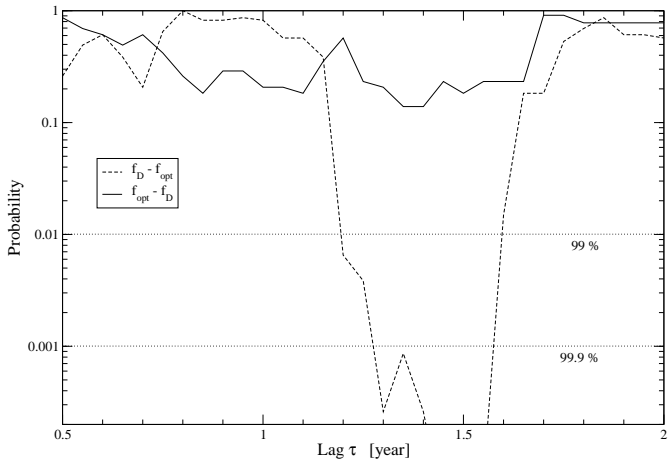}}
  \caption {The significance level of the Spearman's rank correlation
  coefficient for different time delays between $f_{\rm D}$ and
  $f_{\rm opt}$ (dashed line) and $f_{\rm opt}$ and $f_{\rm D}$ (solid
  line) light curves. The horizontal dotted lines correspond to the
  $99~\%$ and $99.9~\%$ confidence levels.}
  \label{D-opt}
\end{figure}

\begin{figure}
\resizebox{\hsize}{!}{\includegraphics{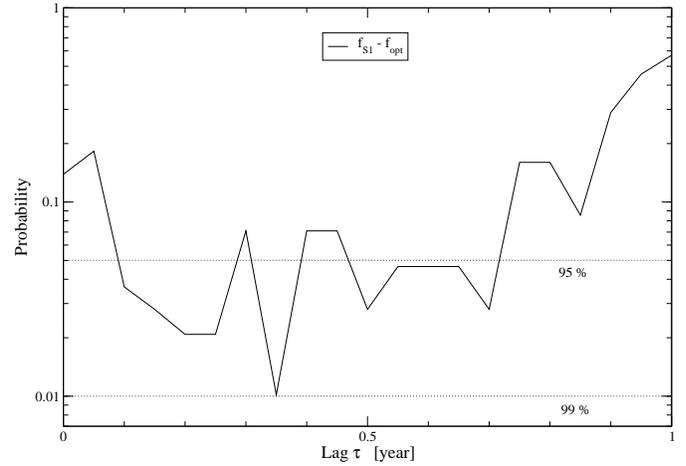}}
  \caption {The significance level of the Spearman's rank correlation
  coefficient for different time delays between $f_{\rm S1}$ and
  $f_{\rm opt}$ light curves. The horizontal dotted lines correspond
  to the $95~\%$ and $99~\%$ confidence levels.  }
  \label{S1-opt}
\end{figure}

A poor radio sampling (10 points) makes it impossible to use the
discrete correlation or cross-correlation functions for determining
the time delay $\tau_{\rm D-S1}$ between $f_{\rm D}$ and $f_{\rm S1}$
(hereafter in the subscripts of $\tau$ the first term indicates the
leading source). Here, we employ the $\chi^2$ minimization method for
linearly interpolated radio light curves. Each light curve is linearly
interpolated, and the data points of one light curve are shifted. For
every value of the shift we calculate the normalized $\chi^2/(N-n)$,
where $N$ is the number of overlapping points and $n=2$ is the number
of fitted parameters.
Application of this method shows that the best time delay is
$\tau_{\rm D-S1}\sim 1$ yr (Fig.~\ref{D-S1}) for both the entire radio
data and for a subsample of eight radio observations made during the
time period of optical measurements (from 1994.6 to 2000.6). The time
delay of $\sim1$ yr is comparable with the mean time $t_{\rm
D-S1}=1.04\pm0.16$ yr at which the moving components C4-C7 (the data
for C8 are too sparse for making reliable estimates) pass the distance
between D and S1. This indicates that the time delay between
variations of radio light curves is related to the passage of moving
components between D and S1 stationary components.

We use a different method to check for correlations between
variability of the jet radio emission and the optical continuum
emission at 5100\AA\ ($f_{\rm opt}$). Using the advantage of a dense
sampling of optical measurements we generate paired radio-optical data
points. For every value of time shift and for every radio measurement
we assign an optical flux which is linearly interpolated between two
nearest epochs of optical observations around the shifted epoch of a
radio measurement. The significance level of the Spearman's rank
correlation coefficient is calculated for different values of the
shift.

\begin{figure}
\resizebox{\hsize}{!}{\includegraphics{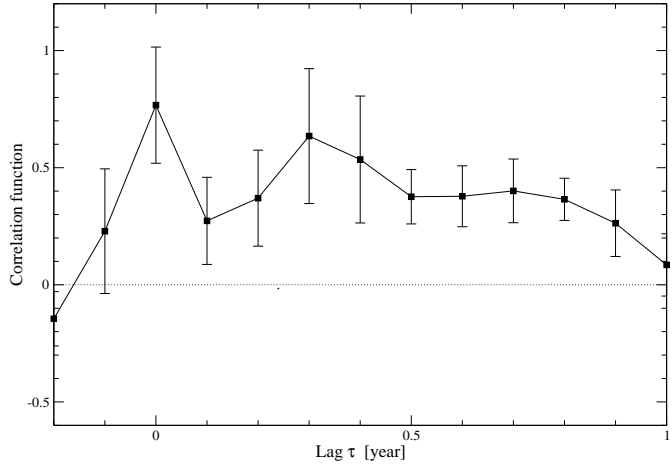}}
  \caption {The discrete correlation function for different lags
  between $f_{\rm S1}$ and $f_{\rm opt}$ during the 1994.5 to 2000
  time period. The 0.1 year bin width of the time delay is adopted to
  minimize the errors of correlation function. 1$\sigma$ error bars
  are presented. }
  \label{dcf}
\end{figure}

This method has been applied to the optical (106 points) and radio
(eight points) data measured between 1992 and 2002. Correlations have
been searched between the optical continuum and radio emission from
the components D, S1, S2, and from a composite flux from D+S1. No
significant correlation exists for the moving features, the stationary
feature S2 and the flux density of D$+$S1 (not shown in
Fig.~\ref{flux-epoch}). A significant correlation ($>99\,\%$, see
Fig.~\ref{D-opt}) is present between $f_{\rm D}$ and $f_{\rm opt}$ for
time lags ranging from $\tau_{\rm D-opt}\sim(1.2-1.6)$ yr. No
significant correlation is found in the $f_{\rm opt} - f_{\rm D}$
relation plane where the optical continuum is assumed to lead the
radio emission (Fig.~\ref{D-opt}). This result strongly suggests that
the radio emission precedes the optical continuum by approximately 1.4
yr, and that there should be a physical relation between variable
radio emission from the feature D of the jet and optical continuum
emission. One should expect that the optical continuum also correlates
with and follows the radio emission of S1 component with time delay of
$\tau_{\rm S1-opt}=\tau_{\rm D-opt}-\tau_{\rm D-S1}\sim 0.4$
yr. Fig.~\ref{S1-opt} shows that there is a positive correlation (at
the $\ga95$~\% confidence level) in the $f_{\rm S1} - f_{\rm opt}$
relation plane for a wide range of time lags from $\sim 0$ yr to $\sim
0.7$ yr. To check this result we use the discrete correlation function
\citep{edelson88} which yields a strong positive correlation between
the radio flux density of S1 (leading source) and the optical
continuum for time lags ranging between 0\,yr and 0.8\,yr with
slightly larger correlation function at small time delays,
$\sim(0-0.4)$\,yr (see Fig.~\ref{dcf}). Both statistical tests show a
positive correlation between variations of $f_{\rm S1}$ and $f_{\rm
opt}$ for $\tau_{\rm S1-opt}\sim(0-0.7)$ yr. It is most probable that
the optical continuum is produced near or at the location of the radio
emission (0-0.4 light year from the feature S1). If the optical
continuum trails the radio emission then the optical and high-energy
photons are likely to be produced by Compton upscattering of radio
photons by the relativistic electrons of the jet \citep{ulrich}.  In
either case, the size of the region from which the optical continuum
is emitted should be very small. This can also be deduced from the
facts that, in 3C\,390.3, the changes of the optical flux at
5100\,\AA\ follow the variations in the UV and soft X-ray emission,
with time delays of about $5 \pm 5$ days \citep{dietrich} and the
overall spectrum in the optical to soft X-ray range is represented by
a single power-law with a spectral index $\alpha=0.89$
\citep{wamsteker}. These observations indicate that a large fraction
of the variable continuum emission is non-thermal and it is produced
in a region of $\sim 5$ light days (0.004 pc) in size.

The link between the optical continuum and the component S1 in the
radio jet is further supported by the correlation between the
continuum light curve and the characteristic epochs $t_\mathrm{D}$ and
$t_\mathrm{S1}$ of the moving components C4--C7. For the components
C4--C7, the epochs $t_\mathrm{S1}$ of separation from the stationary
feature S1 are coincident, within the errors, with the maxima in the
optical continuum (Fig.~\ref{flux-epoch}). All four ejection events
occur within $\sim 0.3$\,yr after a local maximum is reached in the
the intensity of the optical continuum (the average time delay between
the maxima and the epochs $t_\mathrm{S1}$ is $0.18\pm 0.06$
years). The null hypothesis that this happens by chance is rejected at
a confidence level of 99.98\%. This suggests that radio ejection
events of the jet components are coupled with the long-term
variability of optical continuum.

\begin{figure}
\resizebox{\hsize}{!}{\includegraphics{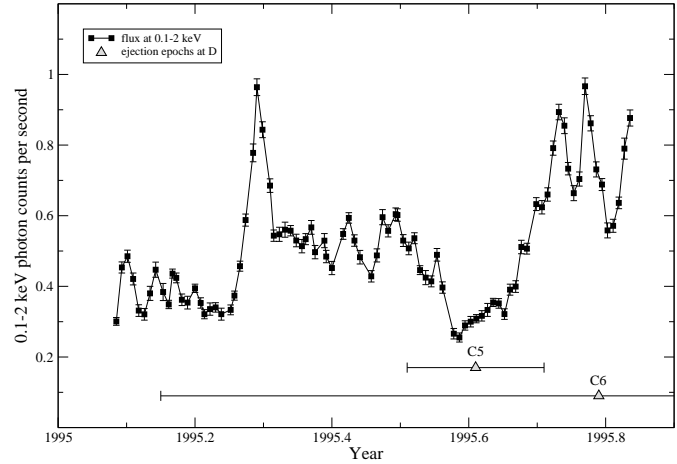}}
  \caption {The variations of soft X-ray fluxes at 0.1-2 keV
  \cite{leighly} for 3C 390.3 between 1995 and 1996 superimposed with
  the epochs of origin of C5 and C6. 1$\sigma$ error bars are
  presented for all data.}
  \label{xray}
\end{figure}

\section{The central sub-parsec-scale region}
The link between the radio emission from D and S1 suggests that the
appearance of these stationary features in the jet is due to physical,
rather than geometrical, factors. The location of stationary features
with respect to the central nucleus is important for understanding the
structure of the central engine and its radiation mechanism producing
the variable continuum emission. If the component D is identified with
the proverbial ``base'' of the counterjet then the component S1 should
be the base of the jet. We should expect that $f_{\rm D}<f_{\rm S1}$
because of relativistic Doppler effect. The fact that $f_{\rm D} \ga
f_{\rm S1}$ during the six year time period (Fig.~\ref{flux-epoch})
rules out the assumption that D is the base of the counterjet. The
component D is likely to be the base of the VLBI jet near the central
engine; the base being the location at which the flow either becomes
supersonic \citep{daly88} or optically thin \citep{konigl81,lobanov98}
or releases the energy contained in the Poynting flux
\citep{romanova96}. The identification of the component D with the
central engine is supported by the correlation between the ejection
epoch, $t_\mathrm{D}$, of the component C5 and the variability of the
X-ray flux at 0.1-2 keV (Fig.~\ref{xray}). The ejection of C5 occurred
after a dip in the X-ray emission and hardening of the spectrum
\citep{leighly} suggesting (similar to \object{3C 120}; Marscher et
al. 2002) that the soft X-ray emitting disk material disappears into
the black hole and a fraction of the infalling matter is ejected into
the jet. The component D should then be associated with the radio
emission coming from the immediate vicinity of the central engine of
3C\,390.3 (assuming that absorption along the line of sight
passing through outer layers of the torus is sufficiently weak). This
emission is probably generated in or above a hot corona at a distance
$\ge$ 200\,$R_\mathrm{s}$ above the accretion disk
\citep{fabian04,ponti04} ($R_\mathrm{s} = 2\,G\,M_\mathrm{bh}/c^2$ is
the Schwarzschild radius for a black hole of mass $M_\mathrm{bh}$,
where $G$ is the Newtonian gravitational constant).

\begin{figure}
\resizebox{\hsize}{!}{\includegraphics{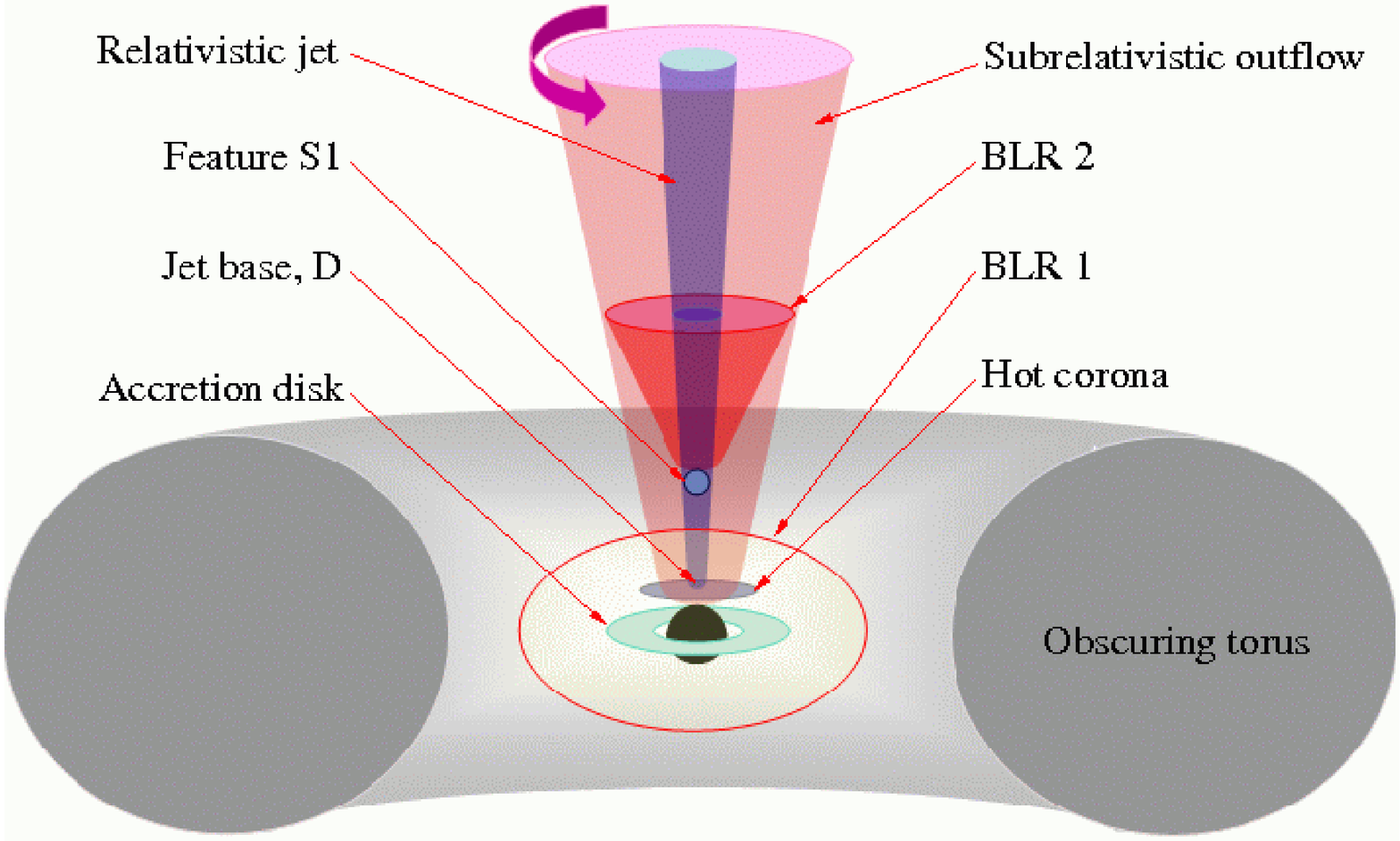}} 
\caption{A sketch of the nuclear region in 3C 390.3. The drawing is
made not to scale and shows only the approaching jet. The broad-line
emission is likely to be generated both near the disk
\citep{peterson02} (BLR\,1, ionized by the emission from a hot corona
\citep{fabian04,ponti04} or the accretion disk; Field \& Rogers 1993) 
and in a rotating subrelativistic outflow \citep{murray97,proga00}
surrounding the jet (BLR\,2, ionized by the emission from the
relativistic plasma in the jet).}
\label{model}
\end{figure}

The component S1 then can be associated with the stationary radio
feature which may be produced by the internal oblique shock formed in
the continuous relativistic flow \citep{gomez95}. In this scenario,
the energy release in the form of variable radiation can be produced
by interaction of relativistically moving compact component or shock
(?) with the \emph{stationary shocks} \citep{gomez95} produced at the
positions of D and S1 of the jet. The time delay between variations of
$f_{\rm D}$ and $f_{\rm S1}$ is equal to the time ($\tau_{\rm
D-S1}=t_{\rm D-S1}$) at which the ejected component passes from D to
S1. This is in fact the case for the 3C\,390.3
(1\,yr\,$\approx1.04$\,yr, see the previous section). The
characteristics of a long-term variability of optical continuum flux
(amplitude of optical flares and their rate) are likely to be related
to the properties of the jet such as the jet ejection rate, structure
and kinematics of the sub-parsec scale jet.

The results presented above imply strongly that the bulk of variable
optical continuum and the broad-line emission associated with it are
generated at a large distance from the central engine (in addition to
the likely contributions from the hidden, ``conventional'' BLR close
to the nucleus).  Adopting an angle\footnote{On the assumption that
the pattern speed and bulk speed of the jet of 3C\,390.3 are
equal, the jet inclination angle $\theta\sim50^{\circ}$, bulk Lorentz
factor $\gamma\sim2$ and beaming angle $\psi \approx
\gamma^{-1}\sim30^{\circ}$ are estimated using the variable Doppler
factor $\delta=1.16$ \citep{lahteenmaki} and the maximum apparent
speed of $1.5\,c$ observed in the compact jet.} of 50$^\circ$ between
the jet and the line of the sight, the corresponding de-projected
distance between D and S1 is $0.3$\,pc$/\sin50^{\circ} \approx
0.4$\,pc.  The optical continuum emission of 3C\,390.3 leads
the double-peaked H$\beta$ line emission by $\sim 30$--$100$ days
\citep{shapo, sergeev}. The localization of the source of optical
continuum with the innermost part of the jet near S1 implies that the
broad line emission originates in the cone within $\sim 100$ light
days at a distance $\ga 0.4$\,pc from the central engine.  The conical
shape of the BLR associated with the outflow arises from a large
fraction of the continuum radiation being beamed into a cone with a
half-opening angle$^2$ $\psi\approx30^{\circ}$.

The structure of the nuclear region implied by the results presented
above is sketched in Fig.~\ref{model}.  In 3C\,390.3, the
changes in the blue and red wings of H$\beta$ emission line occur
quasi-simultaneously \citep{shapo} implying that broad-line region
associated with the component S1 (outflow related BLR\,2 in
Fig.~\ref{model}) is shaping both the blue and red wings (otherwise
one should expect significant time delay between the variability of
the wings).  The observed double-peaked structure of the H$\beta$ line
can be reproduced by a rotation of the approaching jet and a
subrelativistic outflow \citep{murray97,proga00}.  BLR\,2 is evident
in the broad-line emission at times when the jet emission dominates
the optical continuum (see Fig.~\ref{flux-epoch}). BLR\,1 may be
manifested in the broad-line emission around the epochs of minima in
the continuum flux, when the jet contribution to the ionizing
continuum is small.

\section{Discussion and conclusions}
The large distance of the BLR\,2 from the central engine challenges
the existing models in which the broad-line emission is localized
exclusively around the disk or near the central engine
\citep{peterson02}. It also questions the assumption of virialized
motion in the BLR \citep{kaspi00}, which forms the foundation of the
method for estimating black hole masses from reverberation mapping
\citep{peterson02}. Time delays and profile widths measured during
periods when the jet emission is dominant may not necessarily reflect
the Keplerian motion in the disk, but rather trace the rotation and
outward motion in an outflow. This can result in large errors in
estimates of black hole masses made from monitoring of the broad-line
emission.  In the case of 3C\,390.3, the black hole mass
($2.1\times10^9$ solar masses, $M_{\odot}$) estimated effectively from
the measurements near the maximum in the continuum light curve
\citep{shapo} is significantly larger than the values ($3.5$--$4
\times 10^8\,M_{\odot}$) reported in other works
\citep{wandel99,kaspi00}. This difference is reconciled by considering
the line width and the time delay between the optical continuum and
line fluxes near the minimum of the continuum light curve, which
yields $M_\mathrm{bh} = 3.8\times 10^8\,M_{\odot}$. The possible
existence of an outflow-like region in a number of radio-loud AGN
should be taken into account when estimates of the nuclear mass are
made from the variability of broad emission lines.

While the continuum emission at the base of the counterjet is likely
to be too weak to be detected because of the relativistic dimming and
large opacity in the disk, the optical/UV/X-ray line emissions from
BLR\,2 in the counterjet are free from relativistic effects and have
better chances to penetrate the absorbing medium towards the
observer. Detection of a time lag between correlated variability of
emission lines from the far-side and near-side BLRs would be the most
direct observational evidence for double BLRs ionized by the continuum
radiation from the bases of the jet and counterjet.  If the jets are
intrinsically symmetric then a time delay of $\sim 2$ years is
expected for such variations in 3C\,390.3 (assuming the
distance of $\sim 1$ pc between oppositely positioned BLRs and the jet
viewing angle of 50$^{\circ}$).

The presence of the BLR2 in radio-loud AGN is capable of explaining
some of the spectral characteristics of emission lines. Depending on
the orientation of the jet, the approaching and rotating outflow
material in the BLR2 will imprint prominent signatures on the emission
lines. At small viewing angles of the jet the BLR2 may produce
blueshifted and single-peaked broad emission lines, while non-shifted
and double-peak emission lines \citep{eracleous03} will be observed at
large angles of the BLR2 to the line of sight. Similar characteristics
will have the narrow emission lines \citep{boroson05} ionized in the
approaching subrelativistic outflow by the beamed continuum emission
of the jet.

The principal results of this work and its implications are:
\begin{enumerate}
  \item Analysis of combined radio VLBI, optical/UV and X-ray data
  reveals significant correlations between variable optical continuum
  flux (5100\AA) and radio flux density (15\,GHz) of D and S1
  stationary components of the jet (at the 99\% and 95\% confidence
  levels respectively). The optical emission follows the radio flares
  with the mean $\tau_{\rm D-opt}\sim1.4$ yr and $\tau_{\rm
  S1-opt}\sim0.4$ yr. This finding indicates for the physical link
  between the jet and optical continuum: the variable optical
  continuum emission is located in the innermost part of the jet near
  to component S1 and it is of non-thermal origin. This link is also
  supported by the correlation between the local maxima in the optical
  continuum light curve and the epochs at which the moving components
  of the jet pass the stationary radio feature S1.

  \item These results have important implications for the structure of
  the sub-parsec-scale nuclear region. We suggest that (i) the
  characteristics of the long-term variability of optical continuum
  emission are related to the properties of the sub-parsec scale jet
  (ejection rate of the radio components, its structure and
  kinematics), (ii) the bulk of optical variable emission originates
  in the jet at a distance more than or equal to 0.4\,pc from the
  central engine, and (iii) the continuum radiation of the jet forms
  two BLRs (associated with the jet and counterjet) in the
  subrelativistic outflow around the jet.
\end{enumerate}

\begin{acknowledgements}
We acknowledge valuable discussions with N.G.~Bochkarev. This work was
supported by grants from CONACYT (Mexico), INTAS, FEBR and state
program `Astronomy' (Russia). The National Radio Astronomy Observatory
is a facility of the National Science Foundation operated under
cooperative agreement by Associated Universities, Inc.
\end{acknowledgements}

\bibliographystyle{aa} 


\begin{thebibliography}{}
\bibitem[Alef et al.(1996)]{alef} Alef, W., Wu, S.~Y., Preuss, E.,
  Kellermann, K.~I., Qiu, Y.~H. \ 1996, \aap, 308, 376
\bibitem[Boroson(2005)]{boroson05} Boroson, T.\ 2005, \aj, 130, 381
\bibitem[Chiaberge et al.(1999)]{chiaberge99} Chiaberge, M., 
Capetti, A., \& Celotti, A.\ 1999, \aap, 349, 77
\bibitem[Chiaberge et al.(2002)]{chiaberge02} Chiaberge, M., 
Capetti, A., \& Celotti, A.\ 2002, \aap, 394, 791 
\bibitem[Daly et al.(1988)]{daly88} Daly, R.~A., Marscher, A.~P.\
  1988, \apj, 334, 539
\bibitem[Dietrich et al.(1998)]{dietrich} Dietrich, M. {\it et al.}
  1998, \apjs, 115, 185
\bibitem[Edelson \& Krolik(1998)]{edelson88} Edelson, R.~A., Krolik,
  J.~H.\ 1998, \apj, 333, 646
\bibitem[Eracleous \& Halpern(2003)]{eracleous03} Eracleous, M., \&
  Halpern, J.~P.\ 2003, \apj, 599, 886
\bibitem[Fabian(1999)]{fabian99} Fabian, A.\ 1999,  
  Proc. Natl. Acad. Sci. USA, 96, 4749
\bibitem[Fabian(2004)]{fabian04} Fabian, A.~C.\ 2004, astro-ph/0412224
\bibitem[Ferrari(1998)]{ferrari98} Ferrari, A.\ 1998, \araa, 36, 539
\bibitem[Field \& Rogers(1993)]{field93} Field, G.~B., Rogers, R.~D.\
  1993, \apj, 403, 94
\bibitem[Gomez et al.(1995)]{gomez95} Gomez, J.~L., Marti, 
J.~M.~A., Marscher, A.~P., Ibanez, J.~M.~A., \& Marcaide, J.~M.\ 1995, 
\apjl, 449, L19 
\bibitem[Gomez et al.(1997)]{gomez97} Gomez, J.~L., Marti, J.~M.~A.,
  Marscher, A.~P., Ibanez, J.~M.~A., Alberdi, A.\ 1997, \apj, 482, L33
\bibitem[Hardcastle \& Worrall(2000)]{hardcastle00} Hardcastle, 
M.~J., \& Worrall, D.~M.\ 2000, \mnras, 314, 359
\bibitem[Kaspi et al.(2000)]{kaspi00} Kaspi, S. {\it et al.}\ 2000,
  \apj, 533, 631
\bibitem[Kellermann et al.(2004)]{kellermann04} Kellermann, K.~I. {\it
  et al.}  2004, \apj, 609, 539
\bibitem[K\"onigl(1981)]{konigl81} K\"onigl, A.\ 1981, \apj, 243, 700
\bibitem[L\"ahteenm\"aki \& Valtaoja(1999)]{lahteenmaki}
  L\"ahteenm\"aki, A., Valtaoja, E. \ 1999, \apj, 521, 493
\bibitem[Leighly et al.(1997)]{leighly} Leighly, K.~M. {\it et al.} \
  1997, \apj, 483, 767
\bibitem[Lobanov(1998)]{lobanov98} Lobanov, A.~P.\ 1998, \aap, 330, 79
\bibitem[Marscher et al.(2002)]{marscher02} Marscher, A.P. {\it et
  al.} \ 2002, Nature, 417, 625
\bibitem[Murray et al.(1997)]{murray97} Murray, N., Chiang, G.\ 1997,
  \apj, 474, 91
\bibitem[Mushotzky et al.(1993)]{mushotzky93} Mushotzky, R.~F., Done,
  C., Pounds, K.~A.\ 1993, \araa, 31, 717
\bibitem[Pearson(1997)]{pearson} Pearson, T.~J.\ 1997, in ASP
  Conf. Ser. 180, Synthesis imaging in radio astronomy II, eds.
  G.~B.~Taylor, C.~R.~Carilli, R.~A.~Perley, 335
\bibitem[Peterson et al.(2002)]{peterson02} Peterson, B.~M. 2002,
  Advanced Lectures on The Starburst-AGN Connection, eds Aretxaga, I.,
  Kunth, D. \& M\'ujica, R., Singapore World Scientific, 3
\bibitem[Ponti et al.(2004)]{ponti04} Ponti, G., Cappi, M., Dadina,
  M., Malaguti, G.\ 2004, \aap, 417, 451
\bibitem[Proga et al.(2000)]{proga00} Proga, D., Stone, J.~M., Kallman,
  T.~R.\ 2000, \apj, 543, 686
\bibitem[Romanova \& Lovelace(1996)]{romanova96} Romanova, M.~M.,
  Lovelace, R.~V.~E.\ 1996, A\&AS, 120, 583
\bibitem[Shapovalova et al.(2001)]{shapo} Shapovalova, I.~A. {\it et
  al.} \ 2001, \aap, 376, 775
\bibitem[Sergeev et al.(2002)]{sergeev} Sergeev, S.G., Pronik, V.I.,
  Peterson, B.M., Sergeeva, E.A., Zheng, W. \ 2002, \apj, 576, 660
\bibitem[Ulrich et al.(1997)]{ulrich} Ulrich, M., Maraschi, L., Urry,
  C.~M. \ 1997, \araa, 35, 445
\bibitem[Urry \& Padovani(1995)]{urry95} Urry, C.~M., Padovani, P.\
  1995, \pasp, 107, 803
\bibitem[Wamsteker et al.(1997)]{wamsteker} Wamsteker, W., Ting-gui,
  W., Schartel, N., Vio, R. \ 1997, \mnras, 288, 225
\bibitem[Wandel et al.(1999)]{wandel99} Wandel, A., Peterson, B.~M.,
  Malkan, M.~A.\ 1999, \apj, 526, 579
\bibitem[Worrall(2005)]{worrall} Worrall, D.~M. \ 2005, Multiband
  Approach to AGN, eds Lobanov, A.~P. \& Venturi, T. Memorie della
  Societa Astronomica Italiana, 76, 28
\bibitem[Zheng(1996)]{zheng} Zheng, W. \ 1996, \aj, 111, 1498

\end{thebibliography}
{}

\section*{APPENDIX: Properties of the VLBA images and model fits of 3C 390.3}

Table~\ref{tbl-1} compares the CLEAN component models (denoted ``I'')
and Gaussian model fits (denoted ``M'') for the VLBA data used in the
paper. The columns are: $S_\mathrm{total}$ -- total flux density
[mJy/beam]; $S_\mathrm{peak}$ -- peak flux density [mJy/beam];
$S_\mathrm{min}$ -- minimum flux density [mJy/beam]; $\chi^2$ --
goodness of the fit parameter; $\sigma_\mathrm{rms,uv}$ --
root-mean-square between the observed and model visibilities
[mJy]. Last row presents average ratios between the respective image
and model fit properties. The ratios are close to unity for the total
and peak flux densities. The $\chi^2$ parameter and the visibility
r.m.s. are only slightly larger for the Gaussian model fits, which
indicates that the model fits represent the structure adequately. The
higher maximum negative flux density in the model fits (column
$S_\mathrm{min}$) indicates that that the SNR of the Gaussian fits is
on average 1.5 times lower than that of the VLBI images (due to
increased non-Gaussian shapes of low-brightness regions). This
reduction does not affect the fitted values of the component
parameters (albeit it does increase the parameter errors) as it is
related to extended emission associated with the underlying flow.

\begin{table*}
\caption{Properties of the VLBA images and model fits of 3C 390.3.\label{tbl-1}}
\centering
\begin{tabular}{l|rr|rr|rr|rr|rr}
\hline\hline
 Epoch & 
   \multicolumn{2}{|c|}{$S_\mathrm{total}$} & 
   \multicolumn{2}{|c|}{$S_\mathrm{peak}$} & 
   \multicolumn{2}{|c|}{$S_\mathrm{min}$} & 
   \multicolumn{2}{|c|}{$\chi^2$} &
   \multicolumn{2}{|c}{$\sigma_\mathrm{rms,uv}$} \\
      & \multicolumn{1}{c}{I} & \multicolumn{1}{c|}{M} &
 \multicolumn{1}{c}{I} & \multicolumn{1}{c|}{M} &
 \multicolumn{1}{c}{I} & \multicolumn{1}{c|}{M} &
 \multicolumn{1}{c}{I} & \multicolumn{1}{c|}{M} &
 \multicolumn{1}{c}{I} & \multicolumn{1}{c}{M} \\ 
\hline
1994.67 & 423.9 &422.9 &213.3 &211.1 &$-2.2$  &$-3.4$ &1.175 &1.186 &253.1 &253.8 \\
1995.96 & 463.1 &466.2 &190.8 &190.8 &$-1.6$  &$-2.8$ &0.328 &0.334 &119.2 &120.3 \\
1996.37 & 471.6 &477.6 &192.4 &192.6 &$-1.3$  &$-3.0$ &0.453 &0.475 & 83.4 & 85.5 \\
1996.82 & 389.5 &399.1 &174.2 &174.2 &$-1.5$  &$-3.2$ &1.235 &1.504 & 42.5 & 45.2 \\
1997.19 & 370.0 &377.5 &179.1 &180.4 &$-0.7$  &$-2.2$ &1.533 &1.800 & 39.9 & 43.0 \\
1997.66 & 418.3 &425.2 &267.4 &267.2 &$-1.8$  &$-2.2$ &0.575 &0.580 &152.1 &152.5 \\
1998.21 & 392.4 &389.6 &191.2 &191.4 &$-1.1$  &$-1.3$ &0.429 &0.430 & 94.6 & 94.7 \\
1999.85 & 313.2 &315.8 &216.9 &218.6 &$-1.1$  &$-1.3$ &1.103 &1.149 & 31.0 & 31.4 \\
2001.17 & 274.6 &274.3 &203.6 &204.2 &$-1.1$  &$-1.2$ &1.211 &1.216 & 30.7 & 30.8 \\
2002.13 & 308.8 &303.6 &208.4 &210.3 &$-0.9$  &$-1.2$ &1.020 &1.146 & 26.5 & 27.8 \\ 
$<\mathrm{I/M}>$ & \multicolumn{2}{|c|}{0.994} &
\multicolumn{2}{|c|}{0.998} &
\multicolumn{2}{|c|}{0.664} &
\multicolumn{2}{|c|}{0.943} &
\multicolumn{2}{|c}{0.977}  \\
\hline
\end{tabular}
\end{table*}


\end{document}